\documentclass[aps,prd]{revtex4}
\usepackage{subfigure,graphics} 
\usepackage{flafter}
\usepackage{amsmath}
\begin{document}

\title{
Reducing heterotic $M$-theory to five dimensional supergravity on a manifold with boundary}
\author{Ian G. Moss}
\email{ian.moss@ncl.ac.uk}
\affiliation{School of Mathematics and Statistics, Newcastle University, NE1 7RU, UK}
\author{John T. Omotani}
\email{ppxjto@nottingham.ac.uk}
\author{Paul M. Saffin}
\email{paul.saffin@nottingham.ac.uk}  
\affiliation{School of Physics and Astronomy, University of Nottingham, NG7 2RD, UK}

\date{\today}


\begin{abstract}
This paper constructs the reduction of heterotic $M$-theory in eleven dimensions 
to a supergravity model on a manifold with boundary in five dimensions
using a Calabi-Yau three-fold.
New results are presented for the boundary terms in the action and for the
boundary conditions on the bulk fields. Some general features of dualisation on a 
manifold with boundary are used to explain the origin of some topological terms in
the action. The effect of gaugino condensation on the fermion boundary conditions
leads to a `twist' in the chirality of the gravitino which can provide an uplifting
mechanism in the vacuum energy to cancel the cosmological constant after moduli stabilisation.
\end{abstract}
\pacs{PACS number(s): }

\maketitle
\section{introduction}

Heterotic $M$-theory \cite{Horava:1995qa,Horava:1996ma} is based on the idea that one of
the low-energy limits of $M$-theory can be formulated as 11-dimensional
supergravity with matter fields living on two halves of a 10-dimensional boundary. 
The particle content and interactions of the theory, when reduced to four dimensions,
were first studied in the mid-1990's \cite{Witten:1996mz,banks96}, and have been further
developed in the past few years \cite{Donagi:2004ia,Donagi:2004qk,Anderson:2009ge,Anderson:2009mh}.
There has been considerable progress also in fixing the moduli of the theory whilst retaining small
values of the cosmological constant \cite{Curio:2005ew,Braun:2006th,Ahmed:2009ty,Anderson:2011cz}.  

An interesting feature of heterotic $M$-theory is the existence of a
natural five-dimensional reduction which retains the boundary-bulk structure 
\cite{Lukas:1998tt,Lukas:1998yy}. This version of the theory plays an important role 
in some of the moduli stabilisation mechanisms \cite{Braun:2006th,Ahmed:2009ty}.

It is not widely appreciated that the original formulation of Ho\v{r}ava and Witten contained 
some serious problems, which limited the range of validity of their 11-dimensional theory.
The source of these problems lay in the presence of distributions in the supersymmetry
transformations and in the Bianchi identity of the antisymmetric tensor flux. These
terms prevented construction of a supersymmetric action beyond the leading terms
in an expansion parameter $\kappa_{11}{}^{2/3}$, where $\kappa_{11}$ is the gravitational coupling. 
These problems have now been
resolved by a simple modification to the boundary conditions of the theory,
allowing the supersymmetry transformations and the Bianchi identity to remain
free of distributions and resulting in a low energy theory which is supersymmetric to all 
orders in $\kappa_{11}{}^{2/3}$ \cite{Moss:2003bk,Moss:2004ck,Moss:2005zw}.

In many cases it is possible to make progress starting from the original theory of Harava and Witten, 
since the modified
boundary conditions have little effect on the Yukawa couplings, or on the bosonic sector 
per se, where most of the phenomenology resides. However, the new boundary conditions contain 
fermion bilinears, and one place where these become important is in the presence of a 
gaugino condensate. There are ways to sidestep some of the problems in the original
formulation of heterotic $M$-theory to cover gaugino condensation \cite{Horava:1996vs}, but these 
tricks become unnecessary when working with the correct version of the theory, 
and in some cases the original theory gives incorrect 
potentials in four dimensions\cite{Ahmed:2008jz}.

The purpose of the present paper is to re-examine the five-dimensional limit of heterotic $M$-theory
reduced on a Calabi-Yau three-fold. The original work on this
reduction by Lukas et al \cite{Lukas:1998yy} started from the bosonic sector 
of an $E_6\times E_8$ theory and constructed
the bulk-fermion sector from the known five-dimensional supergravity models.
This approach cannot be used on the boundary, where the bulk fields appear
in the boundary action, and of course it says nothing about the boundary conditions.
We shall construct the Lagrangians and boundary conditions for the $E_6\times E_8$ theory,
mostly by direct reduction from 11 dimensions, resorting to supersymmetry only when 
the technical complexity becomes too daunting.  

There are two reasons why we are interested in the five-dimensional limit of heterotic
$M$-theory:
\begin{itemize}
\item Starting from supergravity on a manifold with boundary in eleven dimensions gives
us a new theory of supergravity on a manifold with boundary in five dimensions. 
\item When there is a gaugino condensate, the fermion boundary conditions break the
supersymmetry in five dimensions and lead to a mechanism for cancelling the
cosmological constant.
\end{itemize}
Supergravity theories with boundary matter and dimension less than eleven 
have been constructed only recently. The most detailed models are those in three
dimensions \cite{Belyaev:2007bg}. Examples in five \cite{Belyaev:2005rt} and 
seven dimensions \cite{Pugh:2010ii} have also been constructed,
but the five-dimensional model included distributions, which is something we
are trying to avoid.
The $E_6\times E_8$ reduction of heterotic $M$ theory leads to a new supergravity model
in five dimensions  with 
Yang-Mills and chiral matter multiplets on the boundary.
We omit most of the terms with four fermion fields due to technical complexity, 
but we will include the four-fermion terms that are relevant to gaugino condensation. 
The supersymmetry of the theory is inherited from eleven dimensions, apart from one term
in the supersymmetry variation which cancels the supergravity anomaly in eleven dimensions. 
There is no supersymmetry anomaly in five dimensions, but we can isolate the offending 
supersymmetry variation and regard it as a `higher order' correction. 
 
The gaugino condensate leads to a small chiral `twist' between the fermions on the two
separate boundary components. This creates a five dimensional vacuum energy due to quantum
effects \cite{fabinger00}. Since moduli stabilisation leads to a negative vacuum energy, the quantum
vacuum energy can be used as an `uplifting' mechanism to obtain a small positive 
cosmological constant in four dimensions. 
Previous work has considered this effect in the case of the gravitino in five dimensions 
\cite{Ahmed:2009ty}. We shall show that the other fermion fields are not twisted by the condensate, 
and consequently that the gravitino is the main contributor to the vacuum energy.
 
The conventions used follow Weinberg \cite{Weinberg:2000cr}. The metric signature is $-+\dots+$.
The gamma matrices satisfy $\{\Gamma_I,\Gamma_J\}=2g_{IJ}$ and 
$\Gamma^{I\dots K}=\Gamma^{[I}\dots\Gamma^{K]}$. Dirac conjugates are
$\bar\psi=i\psi^\dagger\Gamma^0$. Index conventions are given in table \ref{table0}.
 
\begin{table}[htb]
\caption{\label{table0}Index conventions. 
The shortage of letters has meant that some indices used in section 2 have had to be re-assigned
in section 3.}
\begin{ruledtabular}
\begin{tabular}{lll}
Index&Description&Section\\
\hline
$I,J,K\dots$&11-dimensional coordinates&2\\
$A,B,C\dots$&10-dimensional coordinates&2\\
$N$&outward normal&2\\
$\alpha,\beta,\gamma\dots$&5-dimensional coordinates&3\\
$\mu,\nu,\rho\dots$&4-dimensional coordinates&3\\
$a,b,c\dots$&holomorphic Calabi-Yau coordinates&3\\
$A,B,C\dots$&SU(2) spinor indices&3\\
$p,q,r\dots$&fundamental $E_6$-group representation&3\\
$i,j,k\dots$&$H^{1,1}$ Calabi-Yau moduli&3\\
$I.J,K\dots$&$E_6$ or $E_8$ Lie algebra &3\\
$L,R$&chiral components&3\\
\end{tabular}
\end{ruledtabular}
\end{table}

\section{The theory in eleven dimensions}

We begin with some of the ingredients of the improved version of low-energy
heterotic $M$-theory described in Ref.\ \  \cite{Moss:2004ck}. 
The theory is formulated on a manifold ${\cal M}$ with a boundary consisting of two disconnected
components $\partial{\cal M}_1$ and $\partial{\cal M}_2$ with identical topology. This is sometimes called
a `downstairs' formulation, as opposed to an `upstairs' formulation which is defined
on a covering space. The supergravity multiplet is placed on ${\cal M}$ and Yang-Mills multiplets 
live on the boundary. Branes may be present, but we leave these out in the simplest 
version of the theory.

The eleven-dimensional part of
the action is conventional for supergravity,
\begin{eqnarray}
S_{SG}=&&{1\over 2 \kappa_{11}^2}\int_{\cal M}\left(-R(\Omega)
-\bar\Psi_I\Gamma^{IJK}D_J(\Omega^*)\Psi_K-\frac1{48}
G_{IJKL}G^{IJKL}\right.
\nonumber\\
&&\left.-\frac{1}{96}\left(\bar\Psi_I\Gamma^{IJKLMP}\Psi_P
+12\bar\Psi^J\Gamma^{KL}\Psi^M\right)G^*_{JKLM}
-\frac{1}{6.11!}
\epsilon^{I_1\dots I_{11}}(C\wedge G\wedge G)_{I_1\dots I_{11}}\right)dv,
\label{actionsg}
\end{eqnarray}
where $G$ is the Abelian field strength and $\Omega$ is the tetrad connection. The combination
$G^*=(G+\hat G)/2$, where hats denote the standardised subtraction of gravitino terms to make a
supercovariant expression. 

The boundary terms which make the action supersymmetric are
\begin{equation}
S_0={1\over  \kappa_{11}^2}\int_{\cal\partial M}\left(
K\mp\frac14\bar\Psi_A\Gamma^{AB}\Psi_B+
\frac12\bar\Psi_A\Gamma^A\Psi_N\right)dv,\label{bcsg}
\end{equation}
where $K$ is the extrinsic curvature of the boundary. The fermionic additions to
the extrinsic curvature term were discovered first for three dimensions in Ref.\  \cite{luckock89}. 
We shall take the  upper sign on the boundary component $\partial {\cal M}_1$ and 
the lower sign on the boundary component $\partial {\cal M}_2$. 

There are also boundary terms with the Yang-Mills multiplets. These are scaled, by a 
parameter $\epsilon$, compared to the supergravity terms,
\begin{equation}
S_1=-{\epsilon\over \kappa_{11}^2}\int_{\cal \partial M}dv
\left(\frac14\left({\rm tr}F^2-\frac12{\rm tr}R^2\right)+
\frac12{\rm tr}\bar\chi\Gamma^AD_A(\hat\Omega^{**})\chi
+\frac14\bar\Psi_A\Gamma^{BC}\Gamma^A{\rm tr}{F}^*_{BC}\chi\right),
\label{action1}
\end{equation}
where $F^*=(F+\hat F)/2$ and $\Omega^{**}=(\Omega+\Omega^*)/2$. The original formulation of Ho\v{r}ava
and Witten contained an extra `$\chi\chi\chi\Psi$' term, but it is not present in the new version.
The formulation given in Ref.\  \cite{Moss:2005zw} was only valid to order $R$, and the extension of
the theory to include the $R^2$ term has been reported recently \cite{Moss:2008ng}.
In the present paper we consider a small curvature limit, and include only
the $O(R^2)$ terms needed in (\ref{action1}) for reducing the theory to five dimensions.
In particular, higher-order terms in the gravitino given in Ref.\  \cite{Moss:2008ng} 
have been omitted. 

The specification of the theory is completed by boundary
conditions for the tangential components of the anti-symmetric tensor,
\begin{equation}
C_{ABC}=\mp \epsilon\,\left(\omega^Y_{ABC}-\frac12\omega^L_{ABC}\right)
\mp\frac14\epsilon\,{\rm tr}\bar\chi\Gamma_{ABC}\chi,\label{cbc}
\end{equation}
where $\omega^Y$ and $\omega^L$ are the Yang-Mills and Lorentz Chern-Simons forms. These boundary
conditions replace the modified Bianchi identity in the old formulation. A suggestion along these
lines was made in the original paper of Ho\v{r}ava and Witten \cite{Horava:1996ma}. 
(This boundary condition is determined by anomaly cancellation and supersymmetry.
It apparently places a restriction on the Abelian symmetry, although, if we follow 
\cite{Moss:1996ip,Moss:1994jj}, we find that fixing $C_{ABC}$ 
on the boundary is consistent with the BRST symmetry of the gauge-fixed action and 
perfectly valid.) Anomaly cancellation fixes the
relative coupling of the supergravity and Yang-Mills sectors,
\begin{equation}
\epsilon={1\over 4\pi}\left({\kappa_{11}\over4\pi}\right)^{2/3}.
\end{equation}
Further details of the anomaly cancellation, and additional 
fermionic Green-Schwarz terms, can be found in
Ref.\  \cite{Moss:2005zw}.

Boundary conditions for the gravitino can be obtained by variation of the action,
\begin{equation}
\Gamma^{AB}\left(P_\pm+\epsilon\Gamma P_\mp \right)\Psi_A=
\epsilon J_Y{}^A,\label{gbc}
\end{equation}
where $P_\pm$ are chiral projectors based on the outward-going normals,
$J_Y$ is the Yang-Mills supercurrent and
\begin{equation}
\Gamma=\frac1{96}{\rm tr}(\bar\chi\Gamma_{ABC}\chi)\Gamma^{ABC}.
\end{equation}
Higher-order gravitino terms
have been omitted in accordance with our small-curvatuture approximation.
(Another way to obtain the gravitino boundary condition is to integrate
the Rarita-Schwinger equation across the orbifold fixed points in the `upstairs'
version of the theory \cite{Moss:2004ck}.)
Similar boundary conditions are placed on the supersymmetry parameter $\eta$,
\begin{equation}
\Gamma^{AB}\left(P_\pm+\epsilon\Gamma P_\mp \right)\eta=0.
\end{equation}
Boundary conditions on the extrinsic curvature follow likewise from
variation of the action. The boundary conditions are supersymmetric,
in the sense that the full set of boundary conditions transforms into
itself under supersymmetry transformations (i.e. each boundary condition
is supersymmetric if the other boundary conditions are imposed). It is a remarkable
fact that these are the only set of boundeary conditions which are consistent
with anomaly cancellation and the unmodified supersymmetry transformations \cite{Moss:2008ng}.

Supersymmetry transformations of the action are carried out subject to the gravitino and anti-symmetric tensor field boundary conditions. Without the $O(R^2)$ terms, the resulting action is supersymmetric {\it to all orders} in the parameter $\epsilon$, apart
from one term which cancels the supersymmetry anomaly \cite{Moss:2005zw}. For the gravity 
anomaly to vanish it becomes necessary to add the higher-order curvature terms to the
action, and then supersymmetry has only been established up to a limited order in the curvature 
\cite{Moss:2008ng}.

\section{The five dimensional reduction}

In this section we shall present some of the details of a simple $E_6\times E_8$ reduction of the 
11-dimensional theory to 5 dimensions. As we mentioned earlier, the bosonic sector for this reduction
has been looked at in some detail in the old version of heterotic $M$-theory, 
especially by Lukas et al.\ \cite{Lukas:1998tt}.  
We shall give some new results for the topological and the fermionic terms in the action, 
but the main results are the set of boundary conditions for the 5-dimensional supergravity on a
manifold with boundary.  

The reduction uses a background gauge field on one boundary component 
$\partial{\cal M}_1$, related to the $SU(3)$ holonomy of the Calabi-Yau space used to 
compactify the internal dimensions. This is chosen because it induces a low-energy background
$G$-flux due to the boundary condition (\ref{cbc}). 

The field content of the reduced theory is given
in table \ref{table1}. The single hypermultiplet includes the volume modulus $V$
along with a real and a complex scalar. This hypermultiplet has an $SU_L(2)\times SU_R(2)$ symmetry,
with a $U(1)$ subgroup gauged by the graviphoton. (We shall use notation based on
$SU_L(2)\times SU_R(2)$, rather than the equivalent description using $SU(2)\times Sp(2)$.)
The number of Abelian multiplets and the number of chiral matter multiplets
are fixed by the Hodge numbers of the Calabi-Yau space. For this paper we shall only consider
the $H^{1,1}$ moduli and leave the $H^{1,2}$ moduli for another occasion.

\begin{table}[htb]
\caption{\label{table1}Field content of the reduced theory on the 5-dimensional manifold
${\cal M}$ with boundary components ${\cal \partial M}_1$ and ${\cal \partial M}_2$.}
\begin{ruledtabular}
\begin{tabular}{llll}
Multiplet&Number&Location&Fields\\
\hline
Graviton multiplet&$1$&${\cal M}$&$\{g_{\alpha\beta},{\cal A}_\alpha,\psi^A_{\alpha}\}$\\
Hypermultiplet&$1$&${\cal M}$&$\{V,\sigma,\xi,\zeta^A\}$\\
Abelian multiplets&$h_{1,1}-1$&${\cal M}$&
$\{{\cal A}^{\perp i},b^i,\lambda^{\perp Ai}\}$\\
$E_6$ Yang-Mills multiplet&$1$&${\cal \partial M}_1$&$\{A_\mu,\chi^A\}$\\
$E_6$ Chiral matter multiplets&$h_{1,1}$&${\cal \partial M}_1$&$\{C^{ip},\eta^{Aip}\}$\\
$E_8$ Yang-Mills multiplet&$1$&${\cal \partial M}_2$&$\{A_\mu,\chi^A\}$\\
\end{tabular}
\end{ruledtabular}
\end{table}

The metric in $11$ dimensions is chosen to reproduce the Einstein-Hilbert
action for the metric in five-dimensions,
\begin{equation}
ds^2=V^{-2/3}g_{\alpha\beta}dx^\alpha dx^\beta+
V^{1/3}(g_{a\bar b}dx^adx^{\bar b}+g_{\bar a b}dx^{\bar a}dx^b),
\end{equation}
where $g_{a\bar b}$ is the metric on a Calabi-Yau space of fixed volume $v$ and
$V$ is the modulus field associated with changes to the Calabi-Yau volume. The background gauge
field on the boundary ${\cal \partial M}_1$ induces the background $G$-flux in the bulk, 
\begin{equation}
G_{ab\bar c\bar d}=\sqrt{2}\,\alpha_i\nu^i{}_{ab\bar c\bar d},
\end{equation}
where the $\alpha_i$ are constants and $\nu^i{}_{ab\bar c\bar d}$ are generators of 
the Hodge-cohomology group $H^{2,2}$ of the Calabi-Yau space. 
Calabi-Yau moduli arise from expanding the K\"ahler form 
$\omega_{a\bar b}=i g_{a\bar b}$ in terms of the generators $\omega_{i a\bar b}$
of the Hodge-cohomology group $H^{1,1}$,
\begin{equation}
\omega_{a\bar b}=b^i\omega_{i a\bar b}.
\end{equation}
We use Hodge duality to choose $\omega_i=2G_{ij}*\nu^j$, where $G_{ij}$ is the moduli space metric defined in
Appendix \ref{app}. The moduli are constrained because the volume of $g_{a\bar b}$ is 
held constant,
\begin{equation}
{\cal K}_{ijk} b^i b^j b^k=6v,
\end{equation}
where the tensor ${\cal K}_{ijk}$ is defined in Appendix \ref{app}.

The components of the antisymmetric tensor define additional scalar fields $\xi$ 
and vector fields ${\cal A}_\alpha^i$,
\begin{eqnarray}
C_{abc}&=&{i\over 2}\xi\epsilon_{abc}\label{canz}\\
C_{a\bar b\alpha}&=&\sqrt{2}{\cal A}_\alpha{}^i\omega_{ia\bar b},
\end{eqnarray}
where $\epsilon_{abc}$ is the covariantly constant $3$-form on the Calabu-Yau space, 
normalised with
\begin{equation}
\epsilon_{abc}\epsilon^{abc}=48.
\end{equation}
The $C_{\alpha\beta\gamma}$ components define a scalar field $\sigma$ through dualisation
(see below).

For the Fermi sector, we shall use the following set of $\Gamma$ matrices
\begin{eqnarray}
\Gamma_\alpha&=&V^{-1/3}\gamma_\alpha\otimes\gamma_7,\\
\Gamma_a&=&V^{1/6} 1\otimes\gamma_a,
\end{eqnarray}
where $\gamma_\alpha$ and $\gamma_a$ are sets of Dirac matrices in five and six dimensions
respectively. 
These are chosen so that $\gamma_\mu^*=\gamma_\mu$ for $\mu=1\dots4$, $\gamma_5^*=-\gamma_5$
and $\gamma_7^*=-\gamma_7$. 

The Calabi-Yau space has two covariantly constant spinors $u_A$, with $\gamma_7u_A=\pm1$ for
$A=1,2$. We define $\bar u^A=(u_A)^\dagger$ and normalise so that
\begin{eqnarray}
\bar u^A u_B&=&\delta^A{}_B,\\
\bar u^A \gamma_7u_B&=&\tau^A{}_B,
\end{eqnarray}
where $\tau$ is a diagonal matrix with $\tau^1{}_1=-\tau^2{}_2=1$.

The gravitino reduces to produce a gravitino $\psi_\alpha^A$ and superpartners $\zeta^A$ of the 
volume $V$ and $\lambda^{Ai}$ of the moduli $b^i$,
\begin{eqnarray}
\Psi_\alpha&=&V^{-1/6}\left(\psi_\alpha^A-i\frac{\sqrt{3}}{2}\gamma_\alpha\zeta^A\right)
\otimes u_A,\\
\Psi_a&=&\frac12V^{1/3}\lambda^{Ai}\otimes\omega_{ia\bar b}\gamma^{\bar b}u_A,
\end{eqnarray}
where $\zeta^A=b_i\lambda^{Ai}\sqrt{2}$. These particular combinations give 
the standard fermion kinetic terms in five dimensions.

On the boundary, we take the gauge field with the simplest non-trivial embedding which has a 
non-zero background $\tilde\omega_a$ on the boundary component
$\partial\mathcal{M}_{1}$, breaking the $E_8$ symmetry down to $E_6$,
\begin{eqnarray}
A_{\mu} & =&\begin{cases}
A_{\mu}\in {\rm ad}(E_{6}) & \text{on }\partial\mathcal{M}_{1},\\
A_{\mu}\in {\rm ad}(E_{8}) & \text{on }\partial\mathcal{M}_{2},\end{cases}\\
A_{a} & =&\begin{cases}
\tilde{\omega}_{a}+\omega_{ia}{}^{b}T_{bp}C^{ip} & \text{on }\partial\mathcal{M}_{1},\\
0 & \text{on }\partial\mathcal{M}_{2},\end{cases}
\end{eqnarray}
where the $T_{bp}$ are the $(3,27)$ $E_8$ generators (see Appendix \ref{gti}).

The gaugino reduces to a four dimensional gaugino $\chi$ and the superpartner $\eta$ of the matter
fields,
\begin{eqnarray}
\chi & =&
\begin{cases}
\chi^A\otimes u_A
+\frac{1}{2}\omega_{ia}{}^bT_{bp}\eta^{Aip}\otimes\gamma^{a}u_A +
\frac{1}{2}\omega_{i\bar a b}T^{bp}\eta^{Ai}{}_{p}\otimes\gamma^{\bar a}u_A&
\text{on }\partial\mathcal{M}_{1},\ \chi^{A}\in {\rm Ad}(E_{6}),\\
\\\chi^A\otimes u_A & \text{on }\partial\mathcal{M}_{2},\ \chi^{A}\in {\rm Ad}(E_{8}).\end{cases}
\end{eqnarray}
We shall use the boundary conditions to confirm later that the SU(2) components of the fermion fields
become chiral components on the boundary.

\subsection{Dualisation on a manifold with boundary}\label{domwb}

During the reduction to five dimensions we replace the antisymmetric tensor $C_{\alpha\beta\gamma}$
with a scalar field $\sigma$ using a duality transformation. The subject of dualisation on a manifold 
with boundary is an interesting issue in its own right, and so we consider this subject in a 
slightly wider context below.

Consider a 5-dimensional action for a 4-form field strength tensor $G$ with Lagrangian
\begin{equation}
{\cal L}_0(G)=A\,G_{\alpha\beta\gamma\delta}G^{\alpha\beta\gamma\delta}
+B^{\alpha\beta\gamma\delta}G_{\alpha\beta\gamma\delta}+D,
\end{equation}
where $A$, $B$ and $D$ depend on other fields. This can be dualised by replacing the field
strength $G$ with an arbitrary antisymmetric tensor $g$ and a scalar $\sigma$ with Lagrangian
\begin{equation}
{\cal L}(g,\sigma)=A\,g_{\alpha\beta\gamma\delta}g^{\alpha\beta\gamma\delta}
+B^{\alpha\beta\gamma\delta}g_{\alpha\beta\gamma\delta}+D
-c\epsilon^{\alpha\beta\gamma\delta\epsilon}\sigma\,
\partial_\epsilon g_{\alpha\beta\gamma\delta},
\end{equation}
where $c$ may depend on other fields. Variation with respect to $\sigma$ vanishes when 
the exterior derivative $dg$ of $g$ vanishes, and the field equations are equivalent to 
the ones obtained from ${\cal L}_0$. 
If the supersymmetry transformation of $g$ is chosen to be identical with the the supersymmetry 
transformations of $G$, then the Lagrangian ${\cal L}(g,\sigma)$ varies into $dg$ terms and
the supersymmetry transformation of $\sigma$ can be chosen to make the action supersymmetric.

Now consider a boundary with a boundary condition on the field strength tensor,
\begin{equation}
G_{\mu\nu\rho\sigma}=f_{\mu\nu\rho\sigma},\label{gf}
\end{equation}
where the indices are tangential to the boundary, and $f$ is some predetermined
tensor, possibly depending on other fields. We can impose this boundary condition
with a boundary action
\begin{equation}
{\cal L}_b(g,\sigma)=
-c\epsilon^{\mu\nu\rho\sigma}(g_{\mu\nu\rho\sigma}-f_{\mu\nu\rho\sigma})\sigma.
\end{equation}
Supersymmetry of the action is assured as long as the boundary conditions,
including this one, transform into one another under supersymmetry.

The dual theory is constructed by varying the action with respect to $g$, and then inserting
the $g$ field equation into the action. This gives
\begin{equation}
{\cal L}(\sigma)={6c^2\over A}
\left(\partial_\epsilon\sigma
-\frac{1}{24c}\epsilon_\epsilon{}^{\alpha\beta\gamma\delta}B_{\alpha\beta\gamma\delta}
\right)^2+D.
\end{equation}
Part of the boundary term cancels an integration by parts, and we are left with,
\begin{equation}
{\cal L}_b(\sigma)=c\epsilon^{\mu\nu\rho\sigma}f_{\mu\nu\rho\sigma}\sigma.\label{lbs}
\end{equation}
Variation of the action now gives a boundary condition,
\begin{equation}
\partial_z\sigma-\frac{1}{24c}\epsilon^{\mu\nu\rho\sigma}B_{\mu\nu\rho\sigma}
={A\over 6c}\epsilon^{\mu\nu\rho\sigma}f_{\mu\nu\rho\sigma}
\end{equation}
Furthermore, supersymmetry has been retained at each step, and so the boundary
conditions of the dual theory transform into one another under supersymmetry.

In the case of heterotic $M$-theory, variation of $g$ in ${\cal L}(g,\sigma)$ gives the field equation
\begin{equation}
g_{\alpha\beta\gamma\delta}=
V^{-2}\epsilon_{\alpha\beta\gamma\delta}{}^\epsilon\left(
\partial_\epsilon\sigma-i(\xi\partial_\epsilon\bar\xi-\bar\xi\partial_\epsilon\xi)
-\alpha_i{\cal A}^i+\hbox{ fermi terms}\right).\label{gsol}
\end{equation}
The boundary condition on the field strength $G$ can be obtained from 
the exterior derivative of the boundary condition on $C$, Eq.\ (\ref{cbc}).
For the $SU(3)\times E_6$ reduction, the boundary has two components, 
${\cal \partial M}_1$ with $E_6$ gauge fields, and ${\cal \partial M}_2$ with 
$E_8$ gauge fields. The boundary terms corresponding to
Eq.\ (\ref{lbs}) on ${\cal \partial M}_2$ resulting from the 
dualisation procedure are
\begin{equation}
{\cal L}_b(\sigma)=-{1\over 8\kappa_5^2}\epsilon\sigma
\epsilon^{\mu\nu\rho\sigma}{\rm tr}\left(F_{\mu\nu}F_{\rho\sigma}\right)
-{1\over 48\kappa_5^2}\epsilon\sigma\epsilon^{\mu\nu\rho\sigma}
\partial_\mu (V^{-1}{\rm tr}\bar\chi\gamma_{\nu\rho\sigma}\chi).\label{hmsig}
\end{equation}
The first term describes a coupling between the bulk $\sigma$ field and the
Pontryagin density of the matter fields. 

This is not quite the complete story for heterotic $M$-theory, because there
are also boundary terms left over from the $CGG$ terms in the action when we integrate
by parts to construct a Lagrangian ${\cal L}_0(G)$. These terms, which depend
on ${\cal A}_\mu$ and $\xi$, combine with (\ref{hmsig})
to produce additional surface terms which have been absorbed into a covariant
derivative term ${\cal D}_\mu\sigma$ in Eq.\ (\ref{yma}) of Appendix C.

\subsection{Bulk action}

The contribution to the action from the bulk is given by
\begin{equation}
S_{SG}={1\over 2\kappa_5^2}\int_{\cal M}{\cal L}_{SG},
\end{equation}
where $\kappa_5^2=\kappa_{11}^2/v$.
The five-dimensional Lagrangian was obtained by Lukas et al.\ by a combination of the reduction ansatz
for the bosonic terms and then by a comparison with the known five-dimensional supergravity models.
We have checked most of the two-fermion terms directly using the reduction ansatz given above. 
Compared to earlier work, we find differences with some of the mass terms. 
There is an independent consistency check on the mass terms from requiring that the 
vacuum energy vanishes. The vacuum energy was evaluated in \cite{Moss:2004un}, and this consistency
check is satisfied by the new Lagrangian, but not by the old one. 

The Lagrangian with up to two fermion fields is given in Appendix \ref{lagrange}.
We use the covariant $\sigma$ derivative based upon Eq.\ (\ref{gaol}), 
\begin{equation}
{\cal D}_\alpha\sigma=\partial_\alpha\sigma
-i\left(\xi\partial_\alpha\bar{\xi}-\bar{\xi}\partial_\alpha\xi\right)
-\alpha_i{\cal A}_\alpha^i.\label{dsigma}
\end{equation}
Cross-terms between the derivatives of the hypermultiplet scalars and fermions
have been absorbed into the fermion derivatives,
\begin{eqnarray}
{\cal D}_\alpha\psi^A_\beta&=&(\nabla_\alpha+{\cal A}_\alpha^i{\cal P}_i)
\psi^A_\beta+\omega_L^A{}_{B\alpha}\psi^B_\beta,\label{dpsi}\\
{\cal D}_\alpha\lambda^{Ai}&=&(\nabla_\alpha+{\cal A}_\alpha^i{\cal P}_i)
\lambda^{Ai}+\partial_\alpha b^j\Gamma^i{}_{jk}\lambda^{Ak}
+\omega_L^A{}_{B\alpha}\lambda^B,\\
{\cal D}_\alpha\zeta^A&=&(\nabla_\alpha+{\cal A}_\alpha^i{\cal P}_i)\zeta^A
+\omega_R^A{}_B\zeta^B,\label{dzeta}
\end{eqnarray}
where ${\cal P}_i=-\frac14iV^{-1}\alpha_i\tau$, and the hypermultiplet $SU(2)$ connections
are
\begin{equation}
\omega_L^A{}_{B\alpha}=\frac14\left(
\begin{array}{cc}
iV^{-1}{\cal D}_\alpha\sigma&-4V^{-1/2}\partial_\alpha\xi\\
4V^{-1/2}\partial_\alpha\bar\xi&-iV^{-1}{\cal D}_\alpha\sigma\\
\end{array}
\right),\qquad
\omega_R^A{}_{B\alpha}=\frac34\left(
\begin{array}{cc}
-iV^{-1}{\cal D}_\alpha\sigma&0\\
0&iV^{-1}{\cal D}_\alpha\sigma\\
\end{array}
\right).
\end{equation}
These connections are associated with the hypermultiplet tetrad
\begin{equation}
E^A{}_{B\alpha}=\left(
\begin{array}{cc}
V^{-1}(\partial_\alpha V-i{\cal D}_\alpha\sigma)&2V^{-1/2}\partial_\alpha\xi\\
-2V^{-1/2}\partial_\alpha\bar\xi&V^{-1}(\partial_\alpha V+i{\cal D}_\alpha\sigma)\\
\end{array}
\right).\label{tetrad}
\end{equation}
The tetrad is defined so that the hypermultiplet kinetic terms
can be written as an $SU(2)$ trace, ${\rm tr}(E_\alpha E^{\dagger\alpha})/4$.

\subsection{Boundary action}

The first surface terms we consider are ones obtained from a reduction of the
the supergravity boundary terms (\ref{bcsg}),
\begin{equation}
S_0={1\over \kappa_5^2}\int_{{\cal \partial M}_1}{\cal L}_0({\cal \partial M}_1)
+{1\over \kappa_5^2}\int_{{\cal \partial M}_2}{\cal L}_0({\cal \partial M}_2).
\end{equation}
The Lagrangians are
\begin{eqnarray}
{\cal L}_0({\cal \partial M}_1)&=&K-\frac{\sqrt{2}}{2}V^{-1}\alpha
-\frac{1}{4}\tau^A{}_B\bar{\psi}_{A\mu}\gamma^{\mu\nu}\bar{\psi}_{\nu}^{B}-
\frac14\tau^A{}_B\bar{\zeta}_{A}\zeta^{B}
-\frac{1}{4}\tau^A{}_BG^\perp_{ij}\bar{\lambda}_A^i\lambda^{Bj}+{\cal T},\\
{\cal L}_0({\cal \partial M}_2)&=&K+\frac{\sqrt{2}}{2}V^{-1}\alpha
+\frac{1}{4}\tau^A{}_B\bar{\psi}_{A\mu}\gamma^{\mu\nu}\bar{\psi}_{\nu}^{B}+
\frac14\tau^A{}_B\bar{\zeta}_{A}\zeta^{B}
+\frac{1}{4}\tau^A{}_BG^\perp_{ij}\bar{\lambda}_A^i\lambda^{Bj}+{\cal T},
\end{eqnarray}
where ${\cal T}$ is a fermionic torsion term which depends on gravitino components normal
to the boundary. The torsion term cancels a total derivative of the torsion 
in $R(\Omega)$, and never appears in the field equations. These boundary Lagrangians
are consistent with the $\zeta=\lambda=0$ case in \cite{Belyaev:2005rt}.

The remaining surface terms contain the matter fields and their couplings to
the bulk supergravity fields,
\begin{equation}
S_{YM}={1\over g^2}\int_{{\cal \partial M}_1}{\cal L}_{YM}({\cal \partial M}_1)
+{1\over g^2}\int_{{\cal \partial M}_2}{\cal L}_{YM}({\cal \partial M}_2),
\end{equation}
where $g^2=\kappa_{11}^2/(\epsilon v)$. The Lagrangian for the $E_8$ Yang-Mills multiplet is
\begin{equation}
{\cal L}_{YM}({\cal \partial M}_2)=
-\frac{1}{4}V{F}_{\mu\nu}^{I}{F}^{I\mu\nu}-\frac{1}{2}\bar{\chi}^I_A\gamma^{\mu}D_{\mu}\chi^{IA}
-\bar\psi_{A\mu}\,j^{A\mu}-\bar\zeta_A\, j^A+\Theta^\mu{\cal D}_\mu\sigma,\label{yma2}
\end{equation}
where $I$ labels the $E_8$ Lie algebra basis. 
The bulk fermions couple to fermionic currents $j^{A\mu}$, $j^A$, and $\sigma$ to the topological 
current $\Theta^\mu$ given by
\begin{eqnarray}
j^{A\mu}&=&\frac14V^{1/2}\gamma^{\rho\sigma}\gamma^\mu F^I{}_{\rho\sigma}\chi^{AI}\label{ja1}\\
j^A&=&i\frac{\sqrt{2}}{4}V^{1/2}\gamma^{\rho\sigma}F^I{}_{\rho\sigma}\chi^{AI}\label{ja2}\\
\Theta^\mu&=&\frac{1}{12}\epsilon^{\mu\nu\rho\sigma}\left(
\omega^Y_{\nu\rho\sigma}
+\frac14V^{-1}\bar\chi^I_A\gamma_{\nu\rho\sigma}\chi^{AI}\right)\label{theta2}
\end{eqnarray}
Note that $j^{A\mu}$ is the usual supercurrent for the gauge multiplet.

The Lagrangian for the $E_6$ Yang-Mills multiplet and the matter terms is given in Appendix
\ref{lagrange}. Part of the potential depends on the $D$-term,
\begin{equation}
D^I=2\bar C^i\Lambda^I C_i.
\end{equation}
The remaining part of the potential depends on the superpotential $W$, which
can be determined most simply from examination of the $\eta$-mass terms,
\begin{equation}
W=\frac{2\sqrt{3}}{3}{\cal K}^{-1}{\cal K}_{ijk}d_{pqr}C^{ip}C^{jq}C^{kr}.
\end{equation}
The coupling to the bulk fields again depends on the supercurrent, now given
in Eqs.\ (\ref{fc1}-\ref{tdef}). There are
also bulk field contributions to the derivatives,
\begin{equation}
{\cal D}_\mu C^{ip}=D_\mu C^{ip}+\partial_\mu b^k\Gamma^i{}_{jk}C^{jp},
\label{dC}
\end{equation}
where the moduli-space connection coefficients are given in Eq.\ (\ref{modcon}).

\subsection{Boundary conditions}

The boundary conditions for the 5-dimensional theory can be obtained by reducing
the boundary conditions Eqs. (\ref{cbc}) and (\ref{gbc}), or by variation of the full 
reduced action, including the boundary terms. The fermionic boundary conditions in five
dimensions are expressed in terms of 
chiral projection operators,
\begin{equation}
P_+{}^A{}_B=\left(
\begin{array}{cc}
P_L&0\\
0&P_R\\
\end{array}
\right),\qquad
P_-{}^A{}_B=\left(
\begin{array}{cc}
P_R&0\\
0&P_L\\
\end{array}
\right),
\end{equation}
where $P_L=\frac12(1+\gamma_5)$ and $P_R=\frac12(1-\gamma_5)$. In order to facilitate
reduction to four dimensions, we take $\gamma_5$ to be the $\gamma$-matrix component in the 
direction of the {\it ingoing} unit normal on the $E_6$ boundary component ${\cal \partial M}_1$ and 
along the {\it outgoing} unit normal on the $E_8$ boundary component ${\cal \partial M}_2$.
We also use ${\cal D}_5$ to denote the derivative along the
{\it inward} unit normal on the $E_6$ boundary, and the {\it outward} unit normal
on the $E_8$ boundary, but for the extrinsic curvature we always use the outgoing normals.

If we drop the 3-fermi terms, then the fermion boundary conditions on the bulk fields become
\begin{eqnarray}
P_-{}^A{}_B\psi^B_\mu&=&\epsilon\left(\delta_\mu{}^\nu-\frac14\gamma_\mu\gamma^\nu\right)
\tau^A{}_Bj^B_\nu,\label{fermibc1}\\
P_+{}^A{}_B\lambda^{Bi}&=&-\epsilon\, \tau^A{}_Bj^{Bi},\\
P_+{}^A{}_B\zeta^B&=&-\epsilon\, \tau^A{}_Bj^B,\label{fermibc3}
\end{eqnarray}
where $\epsilon=\kappa_5^2/g^2$. The fermionic currents on the right-hand
side of these equations are given in Eqs.\ (\ref{fc1}-\ref{fc3}) for the $E_6$ boundary
component, and in Eqs.\ (\ref{ja1}-\ref{ja2}) for the $E_8$ boundary component.
For the matter fields,
\begin{equation}
P_-{}^A{}_B\,\chi^{IB}=P_+{}^A{}_B\,\eta^{Bip}=0.
\end{equation}
These enable us to identify the chiral components of the Majorana spinors as
\begin{eqnarray}
\chi_L{}^I\equiv\chi^{1I},&&\chi_R{}^I\equiv-\chi^{2I},\nonumber\\
\eta_L{}^{ip}\equiv \eta^{2ip},&&\eta_R{}^{ip}\equiv \eta^{1ip}.\label{chiralcomp}
\end{eqnarray}
The minus sign appears in $\chi_R$ to ensure that $\chi\equiv\chi_R+\chi_L$ is real.
Chirality is defined for a conjugate spinor $\bar\chi$ by $\bar\chi_L=\bar\chi P_L$,
so we have $\bar\chi_L=\overline{\chi_R}$.

The boundary conditions on the bosonic fields which result from a reduction of the
boundary condition (\ref{cbc}) on the $E_8$ component ${\cal \partial M}_2$ are
\begin{eqnarray}
\xi&=&-\frac12\epsilon V^{1/2}\bar\chi_L{}^I\chi_L{}^I,\\
{\cal A}^i_\mu&=&\frac{i}{4}\epsilon \tau^A{}_B\,b^i\bar\chi_A{}^I\gamma_\mu\chi^{BI},
\end{eqnarray}
where the moduli-space tensors are defined in Appendix \ref{app}.
These boundary conditions raise a problem with the supersymmetry 
transformation of the ${\cal A}{\cal F}{\cal F}$ term in the action, which
gives a boundary term involving ${\cal A}$. In eleven dimensions, this is the term 
which cancels with 
the supersymmetry anomaly. In five dimensions, it seems that we have to live with
this variation since we have lost the quantum anomaly when we threw out the
high-energy modes. This variation is $O(\epsilon^3)$, and since the string corrections 
in heterotic string theory appear at $O(\epsilon)$, we might regard the $O(\epsilon^3)$ 
variation as a `small' correction.
(In related work on five-dimensional supergravity, Ref.\  \cite{Belyaev:2005rt} 
imposes ${\cal A}^i_\mu=0$, but that paper includes distributions, 
which we seek to avoid.)  

The boundary condition for $\sigma$ on 
${\cal \partial M}_2$ is most easily 
obtained by the variation of $\sigma$ in the full action, including the surface terms
with Lagrangian (\ref{yma2}),
\begin{eqnarray}
{\cal D}_5\sigma&=&-2\epsilon V^2\partial_\mu\Theta^\mu+\frac34iV\tau^A{}_B\,
\bar\zeta_A\gamma_5\zeta^B
-\frac14 i V\tau^A{}_B\,G^\perp_{ij}\bar\lambda_A^i\gamma_5\lambda^{Bj}\nonumber\\
&&+\frac14iV\tau^A{}_B\bar\psi_{A\alpha}\gamma^{\alpha\beta}\gamma_5
\psi^B_\beta
-\frac{\sqrt{2}}{2}V\tau^A{}_B\bar\zeta_A\gamma^\alpha\gamma_5\psi^B_\alpha,
\end{eqnarray}
where $\Theta^\mu$ is given by Eq.\ (\ref{theta2}). From the variation of $V$,
\begin{equation}
\partial_5 V=-\sqrt{2}\alpha+
2\epsilon V^2{\partial{\cal L}_{YM}\over\partial V}
+i{\sqrt{2}\over 2}V\bar\zeta_A\gamma^\alpha\gamma_5\psi^A_\alpha.
\end{equation}
The boundary conditions on the extrinsic curvature
can be obtained by variation of the metric, 
\begin{eqnarray}
K^{\mu\nu}-Kg^{\mu\nu}&=&
\kappa_5^2\,T_{YM}^{\mu\nu}+\frac{\sqrt{2}}{2}\alpha V^{-1}g^{\mu\nu}
+\frac12\tau^A{}_B\bar\psi_{A\rho}\gamma^{\mu\rho}\psi^{B\nu},\nonumber\\
&&+\frac14\tau^A{}_B\left(\bar\psi_{A\rho}\gamma^{\rho\sigma}\psi^B_\sigma 
+\bar\zeta_A\zeta^B
+G^\perp_{ij}\bar\lambda_A^i\lambda^{Bj}\right)g^{\mu\nu},
\end{eqnarray}
where the surface stress tensor for the matter on ${\cal\partial M}_2$ is
\begin{equation}
T_{YM}^{\mu\nu}={1\over g^2}\left(
2{\delta {\cal L}_{YM}\over\delta g_{\mu\nu}}-g^{\mu\nu}{\cal L}_{YM}\right).
\end{equation}
The bulk-fermion bilinear terms in these bosonic boundary conditions can be 
re-written in a variety of ways using the fermion boundary conditions.

On the $E_6$ boundary component ${\cal \partial M}_1$ we use the Lagrangians given in
Appendix (\ref{lagrange}),
\begin{eqnarray}
\xi&=&\frac12\epsilon V^{1/2}\bar\chi_L{}^I\chi_L{}^I-\frac{\sqrt{3}}{3}\epsilon
\epsilon_{abc}d_{pqr}{\cal K}^{-1}{\cal K}_{ijk}C^{ip}C^{jq}C^{kr},\\
{\cal A}^i_\mu&=&-\frac{i}{4}\epsilon \tau^A{}_B\,b^i\bar\chi_A{}^I\gamma_\mu\chi^{BI}
-i\epsilon\,\Gamma^i{}_{jk}
\left(C^{jp}{\cal D}_\mu\bar C^k{}_p-\bar C^k{}_p{\cal D}_\mu C^{jp}\right),\nonumber\\
&&+\frac{i}{4}\epsilon\left(\Gamma^i{}_{jk}-b^iG_{jk}\right)
\left(\bar\eta_R^j{}_p\gamma_\mu\eta_L^{kp}-\bar\eta_L^{jp}\gamma_\mu\eta_R^{k}{}_p\right),\nonumber\\
{\cal D}_5\sigma&=&-2\epsilon V^2\partial_\mu\Theta^\mu+\frac34iV\tau^A{}_B\,
\bar\zeta_A\gamma_5\zeta^B
-\frac14 i V\tau^A{}_B\,G^\perp_{ij}\bar\lambda_A^i\gamma_5\lambda^{Bj}\nonumber\\
&&+\frac14iV\tau^A{}_B\bar\psi_{A\alpha}\gamma^{\alpha\beta}\gamma_5
\psi^B_\beta
-\frac{\sqrt{2}}{2}V\tau^A{}_B\bar\zeta_A\gamma^\alpha\gamma_5\psi^B_\alpha,\\
\partial_5 V&=&-\sqrt{2}\alpha-
2\epsilon V^2{\partial{\cal L}_{YM}\over\partial V}
+i{\sqrt{2}\over 2}V\bar\zeta_A\gamma^\alpha\gamma_5\psi^A_\alpha,\\
K^{\mu\nu}-Kg^{\mu\nu}&=&
\kappa_5^2\,T_{YM}^{\mu\nu}-\frac{\sqrt{2}}{2}\alpha V^{-1}g^{\mu\nu}
-\frac12\tau^A{}_B\bar\psi_{A\rho}\gamma^{\mu\rho}\psi^{B\nu}\nonumber\\
&&-\frac14\tau^A{}_B\left(\bar\psi_{A\rho}\gamma^{\rho\sigma}\psi^B_\sigma 
+\bar\zeta_A\zeta^B
+G^\perp_{ij}\bar\lambda_A^i\lambda^{Bj}\right)g^{\mu\nu}.
\end{eqnarray}
Note that the boundary conditions have a simple solution where all of the fields vanish apart
from $V=1-\sqrt{2} \alpha x^5$ and $g_{\mu\nu}=V^{1/3}\eta_{\mu\nu}$, where
$\eta_{\mu\nu}$ is the flat Minkowski metric. These also satisfy the field
equations, and form the background for a reduction of the theory to four dimensions.

\subsection{Gaugino condensates}

An important question we have to address is the extent to which a gaugino condensate
on either boundary can affect the boundary conditions, and whether the condensate can break 
the supersymmetry as
a consequence. The direct effect of a gaugino condensate on the fermion boundary conditions
is a special feature of the corrected heterotic $M$-theory, where it appears in the chiral
projection as the term $\Gamma$ in (\ref{gbc}). The consequences for the
gravitino in five dimensions have already been addressed in Ref.\  \cite{Ahmed:2009ty},
where the Calabi-Yau space had a topology with Hodge number $h_{1,1}=1$. We are now
in a position to give
the modified boundary conditions for the general case. 

The reduction continues as before, but this time we introduce the gaugino condensate
on the boundary component ${\cal \partial M}_n$,
\begin{equation}
\langle\bar\chi^I\Gamma_{abc}\chi^I\rangle=\Lambda_n\epsilon_{abc},
\end{equation}
where $\Lambda_n\equiv\Lambda_n(V)$. Other boundary fermion fields will be 
set to zero. The condensate appears in the boundary condition 
for the antisymmetric tensor (\ref{cbc}), producing
an affect on the boundary conditions for the field $\xi$ in five dimensions,
\begin{equation}
\xi=
\begin{cases}\displaystyle
\frac12i\epsilon\Lambda_1-\frac{\sqrt{3}}{3}\epsilon\,
\epsilon_{abc}d_{pqr}{\cal K}^{-1}{\cal K}_{ijk}
C^{ip}C^{jq}C^{kr}&\text{ on }{\cal \partial M}_1,\\
\\\displaystyle
-\frac12i\epsilon\Lambda_2
&\text{ on }{\cal \partial M}_2.\\
\end{cases}
\end{equation}
These boundary conditions imply that the field $\xi$ develops a vacuum expectation value 
determined by $\Lambda_1$ and $\Lambda_2$ 
\cite{Ahmed:2008jz}. This is equivalent to generating a background $G$-flux (see Eq.\ (\ref{canz})), 
and it is
this mechanism in heterotic $M$-theory which produces the condensate-induced superpotential $W_g$ 
in four dimensions. 
The four-dimensional supersymmetry is broken, except for special values of $V$ where the 
super-derivatives of $W_g$ vanish and the
four dimensional supersymmetry is restored. There are no other contributions from the 
fermion condensate to the bosonic boundary conditions.

The fermion boundary conditions (\ref{gbc}) include the $\Gamma$-correction to the chiral
projection operator,
\begin{eqnarray}
\left(P_-+\Gamma_n P_+\right){}^A{}_B\psi^B_\mu&=&0,\\
P_+{}^A{}_B\lambda^{Bi}&=&0,\\
P_+{}^A{}_B\zeta^B&=&0,
\end{eqnarray}
where
\begin{equation}
\Gamma_n=\frac12\epsilon\Lambda_n V^{-1/2}\left(
\begin{array}{cc}
0&1\\
1&0\\
\end{array}
\right).
\end{equation}
The gravitino boundary condition can be regarded as a chiral twist between the two boundaries.
The result given here agrees with Ref.\  \cite{Ahmed:2009ty}. 
The gravitino boundary condition breaks the five-dimensional supersymmetry via quantum
effects, specifically by producing a non-zero vacuum energy. This vacuum energy was calculated in
Ref.\  \cite{Ahmed:2009ty}. There are no condensate contributions to the other fermion boundary 
conditions. Note that there are contributions from the condensate-induced
$G$-flux to the fermion mass terms, but these are related to the superpotential and do not 
contribute to the vacuum energy. (The fermion mass terms were used to calculate the
superpotential in Ref.\  \cite{Ahmed:2008jz}.)

\section{conclusion}

We have completed the rather modest task of reducing the simplest version of heterotic $M$-theory
using a Calabi-Yau three-fold to produce a new supergravity model on a manifold with boundary in five dimensions. Many features of the the theory are familiar from the old literature \cite{Lukas:1998tt,Lukas:1998yy}. The new results are the in the boundary action and the boundary conditions on the bulk fields.
These contain features which are worthy of further investigation.

One example is that the Pontryagin density of the Yang-Mills gauge fields appears in the
boundary condition for the hypermultiplet scalar $\sigma$.
The appearance of topological terms here in the boundary conditions has its origin in the dualisation
of the Abelian gauge field on the manifold with boundary (see Sect.\ \ref{domwb}), and is 
the origin of the Pontryagin density terms in the four-dimensional action. It is possible that
this hypermultiplet scalar plays a role in gaugino condensation, since gaugino condensation is
associated with instanton effects. 

The principal motivation for the reduction has been to incorporate gaugino condensation effects
into the boundary conditions, because it has been argued that these can provide an uplifting
mechanism for the potential in the context of moduli stabilisation. The results here are quite
simple: the gravitino boundary conditions give a twist in the chirality of the gravitino which is
proportional to the magnitude of the gaugino condensate. The only effect of the
gaugino condensate on the bosonic sector is to source a background $G$-flux, which contributes to the moduli fields' superpotential.

The consequences of twisting the fermion boundary conditions on the vacuum energy have been
related elsewhere \cite{Ahmed:2009ty} but, so far, no-one has analysed the effects 
on such 1-loop quantum calculations of the coupling between the bulk and 
boundary modes present in the boundary conditions (\ref{fermibc1}-\ref{fermibc3}).
This coupling may also be relevant to the calculation of anomalies on a manifold 
with boundary.

The present work has been restricted to the $H^{1,1}$ moduli of the Calabi-Yau
space. We hope to be able to complete a similar analysis for the $H^{2,1}$ moduli
if there is sufficient interest. More ambitious yet would be to consider other
reductions of heterotic $M$-theory which are more relevant to particle phenomenology
\cite{Donagi:2004ia,Donagi:2004qk,Anderson:2009ge,Anderson:2009mh}.
The main problem here is the inclusion five-branes, which are required for anomaly cancellation.
A consistent approach, like the one we have adopted, should include the back-reaction
of the matter and curvature of the the five-brane, but this is not possible with the
present understanding of the five-brane. However, if we ignore the content of the five-brane,
it should be possible to combine boundary conditions on the bulk fields at the
boundaries with junction conditions across the five-branes.

\acknowledgments
John Omotani and Paul Saffin receive support from the Science and Technology
Facilities Council.

\appendix

\section{Calabi-Yau moduli spaces}\label{app}

Some of the definitions and results concerning the moduli spaces of metric deformations 
of a Calabi-Yau space (based on Ref.\  \cite{Candelas:1990pi}) used in the body of the paper
have been collected together in this appendix. 
The K\"ahler form $\omega_{a\bar b}$ 
and the metric tensor $g_{a\bar b}$ on the Calabi-Yau space are related by
\begin{equation}
\omega_{a\bar b}=ig_{a\bar b}.
\end{equation}
The volume is given by
\begin{equation}
v=\frac16\int\,\omega\wedge\omega\wedge\omega.\label{volume}
\end{equation}
We choose a fixed set of generators of the cohomology group $H^{1,1}$ and denote these by 
$\omega_{ia\bar b}$. The K\"ahler form can be expanded using real moduli fields $b^i$,
\begin{equation}
\omega_{a\bar b}=b^i\omega_{i a\bar b}.
\end{equation}
Two important tensors on the moduli space are the metric
\begin{equation}
G_{ij}={1\over 2v}\int\,\omega_i\wedge*\omega_j,\label{defG}
\end{equation}
and the intersection tensor,
\begin{equation}
{\cal K}_{ijk}=\int\,\omega_i\wedge\omega_j\wedge\omega_k.\label{defK}
\end{equation}
The metric and its inverse can be used to raise and lower indices in the usual way.
We also use the notation,
\begin{equation}
{\cal K}={\cal K}_{ijk}b^ib^jb^k=6v.
\end{equation}
Useful identities which follow from Eqs.\ (\ref{volume}) and (\ref{defG})  are
\begin{equation}
*\,\omega_i=b_i\,\omega\wedge\omega-\omega\wedge\omega_i,\label{soid}
\end{equation}
and
\begin{equation}
G_{ij}=-\frac12\omega_{ia}{}^b\omega_{jb}{}^a.\label{Gid}
\end{equation}
The first identity can be used with (\ref{defG}) and (\ref{defK}) to obtain
another series of useful formulae,
\begin{eqnarray}
{\cal K}^{-1}{\cal K}_{ijk}b^k&=&\frac23 b_ib_j-\frac13G_{ij},\\
{\cal K}^{-1}{\cal K}_{ijk}b^jb^k&=&\frac23 b_i,\qquad
b_ib^i=\frac32.
\end{eqnarray}
These imply that
\begin{equation}
G_{ij}=-\frac12{\partial \ln{\cal K}\over\partial b^i\partial b^j}.\label{Gbb}
\end{equation}
The tangent space to the moduli space can be decomposed into the direction along $b^i$ and the
perpendicular direction by using a projection tensor,
\begin{equation}
\delta^\perp{}_i{}^j=\delta_i{}^j-\frac23 b_i b^j.
\end{equation}
Tensor components orthogonal to $b^i$ will generally be denoted by the superscript $\perp$.

By differentiating (\ref{Gid}) with respect to $b^i$ we obtain a set of metric-connection coefficients,
\begin{equation}
\Gamma_{ijk}=-\frac{i}{2}\omega_{ia}{}^b\omega_{jb}{}^c\omega_{kc}{}^a.
\end{equation}
The Levi-Civita connection components follow from differentiating (\ref{Gbb}),
\begin{equation}
\Gamma_{i(jk)}=-\frac32{\cal K}^{-1}{\cal K}_{ijk}-3 b_{(i}G^\perp{}_{jk)}.
\label{modcon}
\end{equation}
The corresponding curvature components of the Levi-Civita connection are
\begin{equation}
R_{ijkl}=\Gamma_{m(il)}\Gamma^m{}_{(jk)}-\Gamma_{m(ik)}\Gamma^m{}_{(jl)},
\end{equation}
which work out as
\begin{equation}
R_{ijkl}=\frac94{\cal K}^{-2}{\cal K}_{il}{}^m{\cal K}_{kjm}
-\frac94{\cal K}^{-2}{\cal K}_{ik}{}^m{\cal K}_{ljm}
+\frac12G_{il}G_{kj}-\frac12G_{ik}G_{lj}.
\end{equation}

Finally, we collect together the combinations of the cohomology generators which
arise frequently in the dimensional reduction of the action,
\begin{eqnarray}
\omega_{ia}{}^a&=&2ib_i,\\
\omega_{ia}{}^b\omega_{jb}{}^a&=&-2G_{ij},\\
\omega_{ia}{}^b\omega_{jb}{}^c&=&-i\Gamma^k{}_{ij}\omega_{ka}{}^c,\\
\omega_{i[a}{}^{[c}\omega_{jb]}{}^{d]}&=&-{\textstyle\frac34}i{\cal K}^{-1}{\cal K}_{ij}{}^k
\varepsilon_{abe}\varepsilon^{cdf}\omega_k{}^e{}_f,\\
\omega_{ia}{}^{[d}\omega_{jb}{}^e\omega_{kc}{}^{f]}&=&
-{\textstyle\frac{1}{48}}i{\cal K}^{-1}{\cal K}_{ijk}\varepsilon_{abc}\varepsilon^{def},\\
\omega_{ia}{}^b\omega_{jb}{}^c\omega_{kc}{}^d\omega_{ld}{}^a&=&
2\Gamma^m{}_{ij}\Gamma_{mkl}.
\end{eqnarray}

\section{$E_8$ group theory identities}\label{gti}

This appendix contains the identities used to reduce the $E_8$ group theory down to its subgroup
$SU(3)\times E_6$. Although this work is standard material, it is included to make plain our choices
for the normalisation factors and signs.
We start from reduction of the adjoint representation of $E_8$ into $SU(3)\times E_6$
representations,
\begin{equation}
248=(8,1)\oplus(1,78)\oplus(3,27)\oplus(\bar3,\bar{27}).
\end{equation}
The corresponding $E_8$ generators are
\begin{equation}
S^i,\;X^I,\;T_{ap},\;T^{ap},
\end{equation}
where the $S^i$ generate the SU(3) subgroup, the $X^I$ generate the $E_6$ subgroup, 
and $(T_{ap})^\dagger=T^{ap}$
generate the cosets.
We denote the generators of $SU(3)$ in the fundamental representation by $(\lambda^i)_a{}^b$, and
the generators of $E_6$ in the fundamental representation by $(\Lambda^I)_p{}^q$. (The SU(3) indices
$a$ and $b$ used in this appendix are raised and lowered using complex conjugation, not with the
Calabi-Yau metric).

The generators are normalised so that
\begin{eqnarray}
Tr(S^iS^j)&=&30\delta^{ij},\\
Tr(X^IX^J)&=&30\delta^{IJ},\\
Tr(T^{ap}T_{bq})&=&\delta^a_b\delta^p_q,\\
Tr(\lambda^i\lambda^j)&=&\frac12\delta^{ij},\\
Tr(\Lambda^I\Lambda^J)&=&3\delta^{IJ}.
\end{eqnarray}
This hotchpotch of normalisation constants turns out to be convenient.

We require the $T_{ap}$ to transform as the fundamental of SU(3) and the fundamental of $E_6$, so we
find
\begin{eqnarray}
\left[S^i,T_{ap}\right]&=&\lambda^{i\;\;b}_{\;a}T_{bp},\\
\left[X^I,T_{ap}\right]&=&\Lambda^{I\;\;q}_{\;p}T_{aq}.
\end{eqnarray}
For the remaining commutators,
\begin{eqnarray}
\left[T_{ap},T_{bq}\right]&=&\frac{1}{\sqrt{6}}\hat\epsilon_{abc}d_{pqr}T^{cr},\\
\left[T_{ap},T^{bq}\right]&=&-\frac{1}{30}\delta_a{}^b\Lambda^{I\;\;q}_{\;p}X^I
-\frac{1}{30}\delta_p{}^q\lambda^{i\;\;b}_{\;a}S^i,
\end{eqnarray}
where $\hat\epsilon_{abc}$ is the Levi-Civita tensor and $d_{pqr}$ is an 
$E_6$ symmetric tensor, normalised by
\begin{equation}
\hat\epsilon_{abc}\hat\epsilon^{abc}=6,\qquad d_{prs}d^{qrs}=\delta_p{}^q.
\end{equation}
The trace of a triple product becomes
\begin{equation}
Tr\left(T_{ap}T_{bq}T_{cr}\right)=\frac{1}{\sqrt{24}}\hat\epsilon_{abc}d_{pqr}.
\end{equation}
Finally, outer products of the $SU(3)$ and $E_6$ generators are given by
\begin{eqnarray}
(\lambda^i)^a_{\;c}(\lambda^i)^b_{\;d}&=&
\frac12\left(\delta^a_d\delta^b_c-\frac{1}{3}\delta^a_c\delta^b_d\right),\\
(\Lambda^I)^p_{\;q}(\Lambda^I)^r_{\;s}&=&
\frac16\left(\delta^p_q\delta^r_s+3\delta^p_s\delta^r_q
-30d_{qst}d^{prt}\right).
\end{eqnarray}

\section{Lagrangians}\label{lagrange}

The bulk Lagrangian for the $E_6\times E_8$ reduction of heterotic $M$-theory, 
when truncated at the fermion bilinear terms, is given by,
\begin{eqnarray}
{\cal L}_{SG}&=&-R(\Omega)-\frac{1}{2}V^{-2}\partial^{\alpha}V\partial_{\alpha}V
-\frac12 V^{-2}{\cal D}_\alpha\sigma {\cal D}^\alpha\sigma
-2V^{-1}\partial^\alpha\bar\xi\partial_\alpha\xi
-G^\perp{}_{ij}\partial_{\alpha}b^{i}\partial^{\alpha}b^j
\nonumber\\
&&-\frac14{\cal F}_{i\alpha\beta}{\cal F}^{i\alpha\beta}
-{\cal K}^{-1}{\cal K}_{ijk}\epsilon^{\alpha\beta\gamma\delta\epsilon}
{\cal A}^i_\mu{\cal F}^j_{\beta\gamma}{\cal F}^k_{\delta\epsilon}
-\frac12 V^{-2}\alpha_i\alpha^i\nonumber\\
&&-\bar{\psi}_{A \alpha}\gamma^{\alpha\beta\gamma}{\cal D}_{\beta}\psi_{\gamma}^{A}
-\bar{\zeta}_{A}\gamma^{\beta}{\cal D}_{\beta}\zeta^{A}
-G^{\perp}{}_{ij}\bar{\lambda}_A^i\gamma^{\beta}{\cal D}_{\beta}\lambda^{Aj}\nonumber\\
&&+\frac{i\sqrt{2}}{16}\left(\bar \psi_{A\gamma}\gamma^{\alpha\beta\gamma\delta}\psi^A_\delta
+2\bar\psi_A^\alpha\psi^{A\beta}
-G^\perp{}_{ij}\bar\lambda_A^i\gamma^{\alpha\beta}\lambda^{Aj}
-\bar\zeta_A\gamma^{\alpha\beta}\zeta^A\right)b_k{\cal F}^k_{\alpha\beta}\nonumber\\
&&+\frac{\sqrt{2}}{8}G^\perp{}_{ij}\bar\lambda_A^i\gamma^\alpha\gamma^{\beta\gamma}\psi^A_\alpha
{\cal F}^j_{\beta\gamma}
-\frac{i\sqrt{2}}{16}{\cal K}^\perp{}_{ijk}
\bar\lambda_A^i\gamma^{\alpha\beta}\lambda^{Aj}{\cal F}^k_{\alpha\beta}\nonumber\\
&&-\frac{i}{2}G^\perp{}_{ij}\bar\lambda_A^i\gamma^\alpha\gamma^{\beta}\psi^A_\alpha\partial_\alpha
b^j
+\frac{i\sqrt{2}}{2}\bar\zeta_A\gamma^\alpha\gamma^\beta\psi^B_\alpha E^A{}_{B\beta}\nonumber\\
&&+\frac{\sqrt{2}}{4}V^{-1}\alpha \tau^A{}_B\bar\psi_{A\alpha}\gamma^{\alpha\beta}\psi^B_\beta
+\frac{i\sqrt{2}}{2}V^{-1}\alpha^\perp_i \tau^A{}_B\bar\lambda_A^i\gamma^\alpha\psi^B_\alpha
-2iV^{-1}\alpha \tau^A{}_B\bar\zeta_A\gamma^\alpha\psi^B_\alpha\nonumber\\
&&+\frac{\sqrt{2}}{12}V^{-1}\left(9{\cal K}^{-1}{\cal K}^\perp{}_{ijk}+G^\perp{}_{ij}b_k\right)\alpha^k
\tau^A{}_B\bar\lambda_A^i\lambda^{Bj}
+V^{-1}\alpha^\perp_i\tau^A{}_B\bar\zeta\lambda^{Bi}\nonumber\\
&&+\frac{\sqrt{2}}{4}V^{-1}\alpha\, \tau^A{}_B\bar\zeta_A\zeta^B,\label{bulklag}
\end{eqnarray}
where $\alpha=\alpha_i b^i$, and the derivatives are given by Eqs.\ (\ref{dsigma}-\ref{dzeta}).
The supergravity connection $\Omega$ contains fermionic torsion terms which can be 
determined in the 1.5 order formalism by an independent variation of $\Omega$ 
in the Lagrangian.

The boundary contributions to the supergravity Lagrangian are
\begin{eqnarray}
{\cal L}_0({\cal \partial M}_1)&=&K-\frac{\sqrt{2}}{2}V^{-1}\alpha
-\frac{1}{4}\tau^A{}_B\bar{\psi}_{A\mu}\gamma^{\mu\nu}\bar{\psi}_{\nu}^{B}-
\frac14\tau^A{}_B\bar{\zeta}_{A}\zeta^{B}
-\frac{1}{4}\tau^A{}_BG^\perp_{ij}\bar{\lambda}_A^i\lambda^{Bj}+{\cal T},\\
{\cal L}_0({\cal \partial M}_2)&=&K+\frac{\sqrt{2}}{2}V^{-1}\alpha
+\frac{1}{4}\tau^A{}_B\bar{\psi}_{A\mu}\gamma^{\mu\nu}\bar{\psi}_{\nu}^{B}+
\frac14\tau^A{}_B\bar{\zeta}_{A}\zeta^{B}
+\frac{1}{4}\tau^A{}_BG^\perp_{ij}\bar{\lambda}_A^i\lambda^{Bj}+{\cal T},
\end{eqnarray}
where ${\cal T}$ is a torsion term which cancels a total derivative in $R(\Omega)$.
The matter Lagrangian on the $E_6$ boundary ${\cal \partial M}_1$ is
\begin{eqnarray}
{\cal L}_{YM}({\cal \partial M}_1)&=&
-\frac{1}{4}V{F}_{\mu\nu}^{I}{F}^{I\mu\nu}-
\frac{1}{2}\bar{\chi}^I_A\gamma^{\mu}D_{\mu}\chi^{IA}\nonumber\\
&&-2G_{ij}{\cal D}_{\mu}C^{ip}{\cal D}^{\mu}{\bar{C}}^{j}{}_{p}
-\frac12G_{ij}\bar{\eta}_A^{ip}\gamma^{\mu}{\cal D}_{\mu}\eta^{Aj}{}_{p}\nonumber\\
&&-\frac12V^{-1}G^{ij}\partial_{C^i}W\partial_{\bar C^j}\bar W 
-2V^{-1}\bar C^i\Lambda^I C_i\,\bar C^j\Lambda^I C_j\nonumber\\
&&-2V^{-1/2}G_{ij}
\left(\bar{\chi}^{I}_L\eta_L{}^{i}{}_{q}C^{jp}\Lambda^{Iq}{}_{p}+
\bar{\chi}^{I}_R\eta_R{}^{ip}\bar C^j{}_q\Lambda^{Iq}{}_{p}\right)\nonumber\\
&&+\frac{\sqrt{3}}{2}V^{-1/2}{\cal K}^{-1}{\cal K}_{ijk}\left( d_{pqr} C^{kr}\bar\eta^{ip}_L\eta_L{}^{jq}
+d^{pqr} \bar C^k{}_r\bar\eta_{Rp}^i\eta_R{}^{j}{}_q\right)\nonumber\\
&&-\bar\psi_{A\mu}\,j^{A\mu}-\bar\zeta_A\, j^A-G^\perp_{ij}\bar\lambda_A^i\,j^{Aj}
-\Theta^\mu{\cal D}_\mu\sigma\label{yma}.
\end{eqnarray}
The chiral compoents are $\chi_L=\chi^1$, $\chi_R=-\chi^2$,
$\eta_L{}^{ip}=\eta^{2ip}$ and $\eta_R{}^{ip}=\eta^{1ip}$. The bulk
fields couple to supercurrents and a topological current given by
\begin{eqnarray}
j^{A\mu}&=&\frac14V^{1/2}\gamma^{\rho\sigma}\gamma^\mu F^I{}_{\rho\sigma}\chi^{AI}
+V^{-1/2} \tau^A{}_B\bar C^i\Lambda^I C_i\gamma^\mu\chi^{BI}\nonumber\\
&&+\frac{\sqrt{3}}{2}V^{-1/2}{\cal K}^{-1}{\cal K}_{ijk} d_{pqr}\gamma^\mu
(C^{ip}C^{jq}P_L\eta^{Akr}+\bar C^{ip}\bar C^{jq}P_R\eta^{Akr})\nonumber\\
&&+\gamma^\beta\gamma^\mu (D_\beta C^{ip}P_R\eta^A{}_{ip}+
D_\beta \bar C^{ip}P_L\eta^A{}_{ip}),\label{fc1}\\
j^A&=&i\frac{\sqrt{2}}{4}V^{1/2}\gamma^{\rho\sigma}F^I{}_{\rho\sigma}\chi^{AI}
-i\frac{3\sqrt{2}}{2}V^{-1/2} \tau^A{}_B\bar C^i\Lambda^I C_i\chi^{BI}\nonumber\\
&&-i5\sqrt{6}V^{-1/2}{\cal K}^{-1}{\cal K}_{ijk} d_{pqr}
(C^{ip}C^{jq}P_L\eta^{Akr}+\bar C^{ip}\bar C^{jq}P_R\eta^{Akr}),\label{fc2}\\
j_i^A&=&-iV^{-1/2}(6{\cal K}^{-1}{\cal K}_{ijk}+4G_{ij}b_k)\tau^A{}_B
\bar C^j\Lambda^I C^k\chi^{BI}\nonumber\\
&&+2i(6{\cal K}^{-1}{\cal K}_{ijk}+4G_{ij}b_k)\gamma^\alpha (D_\alpha C^{jp}
P_R\eta^{Ak}{}_p+D_\alpha \bar C^{jp}P_L\eta^{Ak}{}_p)\label{fc3}\\
\Theta^\mu&=&\frac{1}{12}\epsilon^{\mu\nu\rho\sigma}\left(
\omega^Y_{\nu\rho\sigma}
+\frac14V^{-1}\bar\chi^I_A\gamma_{\nu\rho\sigma}\chi^{AI}\right.\nonumber\\
&&\left.+ \frac14V^{-1}G_{ij}(\bar\eta_R^i{}_p\gamma_{\mu\nu\rho}\eta_L^{jp}
+\bar\eta_L^{ip}\gamma_{\mu\nu\rho}\eta_R^j{}_p)\right).\label{tdef}
\end{eqnarray}
On the $E_8$ boundary,
\begin{equation}
{\cal L}_{YM}({\cal \partial M}_2)=
-\frac{1}{4}V{F}_{\mu\nu}^{I}{F}^{I\mu\nu}-\frac{1}{2}\bar{\chi}^I_A\gamma^{\mu}D_{\mu}\chi^{IA}
-\bar\psi_{A\mu}\,j^{A\mu}-\bar\zeta_A\, j^A+\Theta^\mu{\cal D}_\mu\sigma.
\end{equation}
The currents are the same as above, with the matter fields $C^{ip}$ and
$\eta^{Aip}$ set to zero.

\section{Supersymmetry transformations}

We start from the 11-dimensional supersymmetry transformations from Green et al.\ 
\cite{Green:1987mn}. We shall denote the supersymmetry parameter in 
five dimensions by $s^A$. Products of three
or more fermion fields have been dropped. After reduction,
the supersymmetry transformations
of the bulk supergravity fields become
\begin{eqnarray}
\delta e^{\hat\alpha}{}_\alpha&=&\frac12 \bar s_A\gamma^{\hat\alpha}\psi_{\alpha}^A,\\
\delta\psi^A_\alpha&=&{\cal D}_\alpha s^A+
i\frac{\sqrt{2}}{6}\left(\gamma_\alpha{}^{\beta\gamma}
-4\delta_\alpha{}^{\beta\gamma}\right)b_i{\cal F}^i_{\beta\gamma} s^A
+\frac16 V^{-1}\alpha\tau^A{}_B\gamma_\alpha\eta^B,\\
\delta{\cal A}_\alpha&=&\frac{i\sqrt{6}}{4}\bar s_A\psi^A_\alpha.
\end{eqnarray}
The derivative ${\cal D}_\alpha$ is the same as the derivative which acts on the
gravitino in Eq.\ (\ref{dpsi}). For the hypermultiplet,
\begin{eqnarray}
\delta V&=&-i\frac{\sqrt{2}}{2}V\bar s_A\zeta^A,\\
\delta\sigma&=&-\frac{\sqrt{2}}{2}\left(
V\tau^A{}_B\bar s_A\zeta^B-V^{1/2}\bar\xi\bar s_2\zeta^1
-V^{1/2}\xi\bar s_1\zeta^2\right),\\
\delta\xi&=&-i\frac{\sqrt{2}}{2} V^{1/2}\bar s_2\zeta^1\\
\delta\zeta&=&-i\frac{\sqrt{2}}{4} E^A{}_{B\alpha}\gamma^\alpha s^B,
\end{eqnarray}
and for the Abelian multiplets,
\begin{eqnarray}
\delta{\cal A}^{\perp i}&=&0,\\
\delta b^i&=&\frac{i}{2}\bar s_A\lambda^{Ai\perp},\\
\delta\lambda^{\perp i}&=&\frac{i}{2}\partial_\alpha b^i\gamma^\alpha s^A
+\frac{\sqrt{2}}{2}\gamma^{\alpha\beta}{\cal F}^{\perp i}_{\alpha\beta}s^A
-G^{\perp ij}{\cal P}_j{}^A{}_Bs^B.
\end{eqnarray}

On the boundaries, the supersymmetry parameter is a Majorana spinor with chiral components
$s_L=s^1$ and $s_R=-s^2$. The supersymmetry transformations of the fields on the
$E_6$ boundary become
\begin{eqnarray}
\delta A_\mu^I&=&\frac12 V^{-1/2}\bar s_A\gamma_\mu\chi^{AI},\\
\delta\chi^{AI}&=&-\frac14 V^{1/2}F^I_{\mu\nu}\gamma^{\mu\nu} s^A
+iV^{-1/2} G_{ij}\bar C^i\Lambda^I C^j\tau^A{}_B s^B,\\
\delta C^{ip}&=&\frac12\bar s_L\eta_L{}^{ip}-\frac14i\Gamma^i{}_{jk}C^{jp}\bar s_A\lambda^{Ak\perp},\\
\delta\eta_L{}^{ip}&=&-{\cal D}_\mu C^{ip}\gamma^\mu s_R
-\sqrt{3}{\cal K}^{-1}{\cal K}^i{}_{jk}d^{pqr}\bar C^j{}_q\bar C^k{}_r s_L
\end{eqnarray}
where ${\cal D}_\mu C^{ip}$ is defined in Eq.\ (\ref{dC}). These reduce to the standard
supersymmetry transformations of the Yang-Mills and chiral multiplets when $V=1$.

\bibliography{paper.bib}

\begin{thebibliography}{31}
\expandafter\ifx\csname natexlab\endcsname\relax\def\natexlab#1{#1}\fi
\expandafter\ifx\csname bibnamefont\endcsname\relax
  \def\bibnamefont#1{#1}\fi
\expandafter\ifx\csname bibfnamefont\endcsname\relax
  \def\bibfnamefont#1{#1}\fi
\expandafter\ifx\csname citenamefont\endcsname\relax
  \def\citenamefont#1{#1}\fi
\expandafter\ifx\csname url\endcsname\relax
  \def\url#1{\texttt{#1}}\fi
\expandafter\ifx\csname urlprefix\endcsname\relax\def\urlprefix{URL }\fi
\providecommand{\bibinfo}[2]{#2}
\providecommand{\eprint}[2][]{\url{#2}}

\bibitem[{\citenamefont{Horava and Witten}(1996{\natexlab{a}})}]{Horava:1995qa}
\bibinfo{author}{\bibfnamefont{P.}~\bibnamefont{Horava}} \bibnamefont{and}
  \bibinfo{author}{\bibfnamefont{E.}~\bibnamefont{Witten}},
  \bibinfo{journal}{Nucl. Phys.} \textbf{\bibinfo{volume}{B460}},
  \bibinfo{pages}{506} (\bibinfo{year}{1996}{\natexlab{a}}),
  \eprint{hep-th/9510209}.

\bibitem[{\citenamefont{Horava and Witten}(1996{\natexlab{b}})}]{Horava:1996ma}
\bibinfo{author}{\bibfnamefont{P.}~\bibnamefont{Horava}} \bibnamefont{and}
  \bibinfo{author}{\bibfnamefont{E.}~\bibnamefont{Witten}},
  \bibinfo{journal}{Nucl. Phys.} \textbf{\bibinfo{volume}{B475}},
  \bibinfo{pages}{94} (\bibinfo{year}{1996}{\natexlab{b}}),
  \eprint{hep-th/9603142}.

\bibitem[{\citenamefont{Witten}(1996)}]{Witten:1996mz}
\bibinfo{author}{\bibfnamefont{E.}~\bibnamefont{Witten}},
  \bibinfo{journal}{Nucl. Phys.} \textbf{\bibinfo{volume}{B471}},
  \bibinfo{pages}{135} (\bibinfo{year}{1996}), \eprint{hep-th/9602070}.

\bibitem[{\citenamefont{Banks and Dine}(1996)}]{banks96}
\bibinfo{author}{\bibfnamefont{T.}~\bibnamefont{Banks}} \bibnamefont{and}
  \bibinfo{author}{\bibfnamefont{M.}~\bibnamefont{Dine}},
  \bibinfo{journal}{Nucl. Phys.} \textbf{\bibinfo{volume}{B479}},
  \bibinfo{pages}{173} (\bibinfo{year}{1996}), \eprint{hep-th/9605136}.

\bibitem[{\citenamefont{Donagi et~al.}(2004{\natexlab{a}})\citenamefont{Donagi,
  He, Ovrut, and Reinbacher}}]{Donagi:2004ia}
\bibinfo{author}{\bibfnamefont{R.}~\bibnamefont{Donagi}},
  \bibinfo{author}{\bibfnamefont{Y.-H.} \bibnamefont{He}},
  \bibinfo{author}{\bibfnamefont{B.~A.} \bibnamefont{Ovrut}}, \bibnamefont{and}
  \bibinfo{author}{\bibfnamefont{R.}~\bibnamefont{Reinbacher}},
  \bibinfo{journal}{JHEP} \textbf{\bibinfo{volume}{12}}, \bibinfo{pages}{054}
  (\bibinfo{year}{2004}{\natexlab{a}}), \eprint{hep-th/0405014}.

\bibitem[{\citenamefont{Donagi et~al.}(2004{\natexlab{b}})\citenamefont{Donagi,
  He, Ovrut, and Reinbacher}}]{Donagi:2004qk}
\bibinfo{author}{\bibfnamefont{R.}~\bibnamefont{Donagi}},
  \bibinfo{author}{\bibfnamefont{Y.-H.} \bibnamefont{He}},
  \bibinfo{author}{\bibfnamefont{B.~A.} \bibnamefont{Ovrut}}, \bibnamefont{and}
  \bibinfo{author}{\bibfnamefont{R.}~\bibnamefont{Reinbacher}},
  \bibinfo{journal}{Phys. Lett.} \textbf{\bibinfo{volume}{B598}},
  \bibinfo{pages}{279} (\bibinfo{year}{2004}{\natexlab{b}}),
  \eprint{hep-th/0403291}.

\bibitem[{\citenamefont{Anderson
  et~al.}(2010{\natexlab{a}})\citenamefont{Anderson, Gray, Grayson, He, and
  Lukas}}]{Anderson:2009ge}
\bibinfo{author}{\bibfnamefont{L.~B.} \bibnamefont{Anderson}},
  \bibinfo{author}{\bibfnamefont{J.}~\bibnamefont{Gray}},
  \bibinfo{author}{\bibfnamefont{D.}~\bibnamefont{Grayson}},
  \bibinfo{author}{\bibfnamefont{Y.-H.} \bibnamefont{He}}, \bibnamefont{and}
  \bibinfo{author}{\bibfnamefont{A.}~\bibnamefont{Lukas}},
  \bibinfo{journal}{Commun. Math. Phys.} \textbf{\bibinfo{volume}{297}},
  \bibinfo{pages}{95} (\bibinfo{year}{2010}{\natexlab{a}}), \eprint{0904.2186}.

\bibitem[{\citenamefont{Anderson
  et~al.}(2010{\natexlab{b}})\citenamefont{Anderson, Gray, He, and
  Lukas}}]{Anderson:2009mh}
\bibinfo{author}{\bibfnamefont{L.~B.} \bibnamefont{Anderson}},
  \bibinfo{author}{\bibfnamefont{J.}~\bibnamefont{Gray}},
  \bibinfo{author}{\bibfnamefont{Y.-H.} \bibnamefont{He}}, \bibnamefont{and}
  \bibinfo{author}{\bibfnamefont{A.}~\bibnamefont{Lukas}},
  \bibinfo{journal}{JHEP} \textbf{\bibinfo{volume}{02}}, \bibinfo{pages}{054}
  (\bibinfo{year}{2010}{\natexlab{b}}), \eprint{0911.1569}.

\bibitem[{\citenamefont{Ahmed and Moss}(2010)}]{Ahmed:2009ty}
\bibinfo{author}{\bibfnamefont{N.}~\bibnamefont{Ahmed}} \bibnamefont{and}
  \bibinfo{author}{\bibfnamefont{I.~G.} \bibnamefont{Moss}},
  \bibinfo{journal}{Nucl. Phys.} \textbf{\bibinfo{volume}{B833}},
  \bibinfo{pages}{133} (\bibinfo{year}{2010}), \eprint{0907.1602}.

\bibitem[{\citenamefont{Curio et~al.}(2006)\citenamefont{Curio, Krause, and
  Lust}}]{Curio:2005ew}
\bibinfo{author}{\bibfnamefont{G.}~\bibnamefont{Curio}},
  \bibinfo{author}{\bibfnamefont{A.}~\bibnamefont{Krause}}, \bibnamefont{and}
  \bibinfo{author}{\bibfnamefont{D.}~\bibnamefont{Lust}},
  \bibinfo{journal}{Fortsch. Phys.} \textbf{\bibinfo{volume}{54}},
  \bibinfo{pages}{225} (\bibinfo{year}{2006}), \eprint{hep-th/0502168}.

\bibitem[{\citenamefont{Braun and Ovrut}(2006)}]{Braun:2006th}
\bibinfo{author}{\bibfnamefont{V.}~\bibnamefont{Braun}} \bibnamefont{and}
  \bibinfo{author}{\bibfnamefont{B.~A.} \bibnamefont{Ovrut}},
  \bibinfo{journal}{JHEP} \textbf{\bibinfo{volume}{07}}, \bibinfo{pages}{035}
  (\bibinfo{year}{2006}), \eprint{hep-th/0603088}.

\bibitem[{\citenamefont{Anderson et~al.}(2011)\citenamefont{Anderson, Gray,
  Lukas, and Ovrut}}]{Anderson:2011cz}
\bibinfo{author}{\bibfnamefont{L.~B.} \bibnamefont{Anderson}},
  \bibinfo{author}{\bibfnamefont{J.}~\bibnamefont{Gray}},
  \bibinfo{author}{\bibfnamefont{A.}~\bibnamefont{Lukas}}, \bibnamefont{and}
  \bibinfo{author}{\bibfnamefont{B.}~\bibnamefont{Ovrut}},
  \bibinfo{journal}{Phys. Rev.} \textbf{\bibinfo{volume}{D83}},
  \bibinfo{pages}{106011} (\bibinfo{year}{2011}), \eprint{1102.0011}.

\bibitem[{\citenamefont{Lukas et~al.}(1999{\natexlab{a}})\citenamefont{Lukas,
  Ovrut, Stelle, and Waldram}}]{Lukas:1998tt}
\bibinfo{author}{\bibfnamefont{A.}~\bibnamefont{Lukas}},
  \bibinfo{author}{\bibfnamefont{B.~A.} \bibnamefont{Ovrut}},
  \bibinfo{author}{\bibfnamefont{K.~S.} \bibnamefont{Stelle}},
  \bibnamefont{and} \bibinfo{author}{\bibfnamefont{D.}~\bibnamefont{Waldram}},
  \bibinfo{journal}{Nucl. Phys.} \textbf{\bibinfo{volume}{B552}},
  \bibinfo{pages}{246} (\bibinfo{year}{1999}{\natexlab{a}}),
  \eprint{hep-th/9806051}.

\bibitem[{\citenamefont{Lukas et~al.}(1999{\natexlab{b}})\citenamefont{Lukas,
  Ovrut, Stelle, and Waldram}}]{Lukas:1998yy}
\bibinfo{author}{\bibfnamefont{A.}~\bibnamefont{Lukas}},
  \bibinfo{author}{\bibfnamefont{B.~A.} \bibnamefont{Ovrut}},
  \bibinfo{author}{\bibfnamefont{K.~S.} \bibnamefont{Stelle}},
  \bibnamefont{and} \bibinfo{author}{\bibfnamefont{D.}~\bibnamefont{Waldram}},
  \bibinfo{journal}{Phys. Rev.} \textbf{\bibinfo{volume}{D59}},
  \bibinfo{pages}{086001} (\bibinfo{year}{1999}{\natexlab{b}}),
  \eprint{hep-th/9803235}.

\bibitem[{\citenamefont{Moss}(2003)}]{Moss:2003bk}
\bibinfo{author}{\bibfnamefont{I.~G.} \bibnamefont{Moss}},
  \bibinfo{journal}{Phys. Lett.} \textbf{\bibinfo{volume}{B577}},
  \bibinfo{pages}{71} (\bibinfo{year}{2003}), \eprint{hep-th/0308159}.

\bibitem[{\citenamefont{Moss}(2005)}]{Moss:2004ck}
\bibinfo{author}{\bibfnamefont{I.~G.} \bibnamefont{Moss}},
  \bibinfo{journal}{Nucl. Phys.} \textbf{\bibinfo{volume}{B729}},
  \bibinfo{pages}{179} (\bibinfo{year}{2005}), \eprint{hep-th/0403106}.

\bibitem[{\citenamefont{Moss}(2006)}]{Moss:2005zw}
\bibinfo{author}{\bibfnamefont{I.~G.} \bibnamefont{Moss}},
  \bibinfo{journal}{Phys. Lett.} \textbf{\bibinfo{volume}{B637}},
  \bibinfo{pages}{93} (\bibinfo{year}{2006}), \eprint{hep-th/0508227}.

\bibitem[{\citenamefont{Horava}(1996)}]{Horava:1996vs}
\bibinfo{author}{\bibfnamefont{P.}~\bibnamefont{Horava}},
  \bibinfo{journal}{Phys. Rev.} \textbf{\bibinfo{volume}{D54}},
  \bibinfo{pages}{7561} (\bibinfo{year}{1996}), \eprint{hep-th/9608019}.

\bibitem[{\citenamefont{Ahmed and Moss}(2008)}]{Ahmed:2008jz}
\bibinfo{author}{\bibfnamefont{N.}~\bibnamefont{Ahmed}} \bibnamefont{and}
  \bibinfo{author}{\bibfnamefont{I.~G.} \bibnamefont{Moss}},
  \bibinfo{journal}{JHEP} \textbf{\bibinfo{volume}{12}}, \bibinfo{pages}{108}
  (\bibinfo{year}{2008}), \eprint{0809.2244}.

\bibitem[{\citenamefont{Belyaev and van Nieuwenhuizen}(2008)}]{Belyaev:2007bg}
\bibinfo{author}{\bibfnamefont{D.~V.} \bibnamefont{Belyaev}} \bibnamefont{and}
  \bibinfo{author}{\bibfnamefont{P.}~\bibnamefont{van Nieuwenhuizen}},
  \bibinfo{journal}{JHEP} \textbf{\bibinfo{volume}{02}}, \bibinfo{pages}{047}
  (\bibinfo{year}{2008}), \eprint{0711.2272}.

\bibitem[{\citenamefont{Belyaev}(2006)}]{Belyaev:2005rt}
\bibinfo{author}{\bibfnamefont{D.~V.} \bibnamefont{Belyaev}},
  \bibinfo{journal}{JHEP} \textbf{\bibinfo{volume}{01}}, \bibinfo{pages}{047}
  (\bibinfo{year}{2006}), \eprint{hep-th/0509172}.

\bibitem[{\citenamefont{Pugh et~al.}(2011)\citenamefont{Pugh, Sezgin, and
  Stelle}}]{Pugh:2010ii}
\bibinfo{author}{\bibfnamefont{T.~G.} \bibnamefont{Pugh}},
  \bibinfo{author}{\bibfnamefont{E.}~\bibnamefont{Sezgin}}, \bibnamefont{and}
  \bibinfo{author}{\bibfnamefont{K.~S.} \bibnamefont{Stelle}},
  \bibinfo{journal}{JHEP} \textbf{\bibinfo{volume}{02}}, \bibinfo{pages}{115}
  (\bibinfo{year}{2011}), \eprint{1008.0726}.

\bibitem[{\citenamefont{Fabinger and Horava}(2000)}]{fabinger00}
\bibinfo{author}{\bibfnamefont{M.}~\bibnamefont{Fabinger}} \bibnamefont{and}
  \bibinfo{author}{\bibfnamefont{P.}~\bibnamefont{Horava}}
  (\bibinfo{year}{2000}), \bibinfo{note}{hep-th/0002073}.

\bibitem[{\citenamefont{Weinberg}(2002)}]{Weinberg:2000cr}
\bibinfo{author}{\bibfnamefont{S.}~\bibnamefont{Weinberg}},
  \emph{\bibinfo{title}{{The quantum theory of fields. Vol. 3: Supersymmetry}}}
  (\bibinfo{publisher}{Cambridge, UK: Univ. Pr.}, \bibinfo{year}{2002}).

\bibitem[{\citenamefont{Luckock and Moss}(1989)}]{luckock89}
\bibinfo{author}{\bibfnamefont{H.~C.} \bibnamefont{Luckock}} \bibnamefont{and}
  \bibinfo{author}{\bibfnamefont{I.~G.} \bibnamefont{Moss}},
  \bibinfo{journal}{Class Quantum grav} \textbf{\bibinfo{volume}{6}},
  \bibinfo{pages}{1993} (\bibinfo{year}{1989}).

\bibitem[{\citenamefont{Moss}(2008)}]{Moss:2008ng}
\bibinfo{author}{\bibfnamefont{I.~G.} \bibnamefont{Moss}},
  \bibinfo{journal}{JHEP} \textbf{\bibinfo{volume}{11}}, \bibinfo{pages}{067}
  (\bibinfo{year}{2008}), \eprint{0810.1662}.

\bibitem[{\citenamefont{Moss and Silva}(1997)}]{Moss:1996ip}
\bibinfo{author}{\bibfnamefont{I.~G.} \bibnamefont{Moss}} \bibnamefont{and}
  \bibinfo{author}{\bibfnamefont{P.~J.} \bibnamefont{Silva}},
  \bibinfo{journal}{Phys. Rev.} \textbf{\bibinfo{volume}{D55}},
  \bibinfo{pages}{1072} (\bibinfo{year}{1997}), \eprint{gr-qc/9610023}.

\bibitem[{\citenamefont{Moss and Poletti}(1994)}]{Moss:1994jj}
\bibinfo{author}{\bibfnamefont{I.~G.} \bibnamefont{Moss}} \bibnamefont{and}
  \bibinfo{author}{\bibfnamefont{S.~J.} \bibnamefont{Poletti}},
  \bibinfo{journal}{Phys. Lett.} \textbf{\bibinfo{volume}{B333}},
  \bibinfo{pages}{326} (\bibinfo{year}{1994}), \eprint{gr-qc/9405044}.

\bibitem[{\citenamefont{Moss and Norman}(2004)}]{Moss:2004un}
\bibinfo{author}{\bibfnamefont{I.~G.} \bibnamefont{Moss}} \bibnamefont{and}
  \bibinfo{author}{\bibfnamefont{J.~P.} \bibnamefont{Norman}},
  \bibinfo{journal}{JHEP} \textbf{\bibinfo{volume}{09}}, \bibinfo{pages}{020}
  (\bibinfo{year}{2004}), \eprint{hep-th/0401181}.

\bibitem[{\citenamefont{Candelas and de~la Ossa}(1991)}]{Candelas:1990pi}
\bibinfo{author}{\bibfnamefont{P.}~\bibnamefont{Candelas}} \bibnamefont{and}
  \bibinfo{author}{\bibfnamefont{X.}~\bibnamefont{de~la Ossa}},
  \bibinfo{journal}{Nucl. Phys.} \textbf{\bibinfo{volume}{B355}},
  \bibinfo{pages}{455} (\bibinfo{year}{1991}).

\bibitem[{\citenamefont{Green et~al.}(1987)\citenamefont{Green, Schwarz, and
  Witten}}]{Green:1987mn}
\bibinfo{author}{\bibfnamefont{M.~B.} \bibnamefont{Green}},
  \bibinfo{author}{\bibfnamefont{J.~H.} \bibnamefont{Schwarz}},
  \bibnamefont{and} \bibinfo{author}{\bibfnamefont{E.}~\bibnamefont{Witten}},
  \emph{\bibinfo{title}{{Superstring theory. Vol. 2: Loop amplitudes, anomalies
  and phenomenology}}} (\bibinfo{publisher}{Cambridge, UK: Univ. Pr.},
  \bibinfo{year}{1987}), \bibinfo{note}{( Cambridge Monographs On Mathematical
  Physics)}.

\end{thebibliography}

\end{document}